\documentclass[journal=jacsat,manuscript=article,
layout=onecolumn]{achemso}
\usepackage[version=3]{mhchem}
\usepackage{lineno}
\usepackage{xr-hyper}
\usepackage{hyperref}
\usepackage{xcolor}
\usepackage{xr}
\modulolinenumbers[5]
\usepackage[numbers]{natbib}
\usepackage{graphicx}
\usepackage{dcolumn}
\usepackage{float}
\usepackage{bm}
\usepackage{csquotes}
\usepackage{amsmath,amssymb}
\usepackage{mathrsfs}
\usepackage{gensymb}
\usepackage{epstopdf}

\author{Snehal Mandal}
\email{snhlmandal@gmail.com}
\affiliation[SINP1]
{Saha Institute of Nuclear Physics, A CI of Homi Bhabha National Institute, 1/AF, Bidhannagar, Kolkata 700064, India}
\altaffiliation{Current address: Centre for Quantum Engineering, Research and Education, TCG Centres for Research and Education in Science and Technology (TCG CREST), EM Block, Sector V, Salt Lake, Kolkata 700091, India}
\author{Sandip Halder}
\affiliation[SINP1]
{Saha Institute of Nuclear Physics, A CI of Homi Bhabha National Institute, 1/AF, Bidhannagar, Kolkata 700064, India}
\author{Biswarup Satpati}
\affiliation[SINP1]
{Saha Institute of Nuclear Physics, A CI of Homi Bhabha National Institute, 1/AF, Bidhannagar, Kolkata 700064, India}
\author{Kalpataru Pradhan}
\email{kalpataru.pradhan@saha.ac.in}
\affiliation[SINP1]
{Saha Institute of Nuclear Physics, A CI of Homi Bhabha National Institute, 1/AF, Bidhannagar, Kolkata 700064, India}
\author{I. Das}
\affiliation[SINP1]
{Saha Institute of Nuclear Physics, A CI of Homi Bhabha National Institute, 1/AF, Bidhannagar, Kolkata 700064, India}
\title[] {Tailoring the interfacial magnetic interaction in epitaxial La$_{0.7}$Sr$_{0.3}$MnO$_3$/Sm$_{0.5}$Ca$_{0.5}$MnO$_3$ heterostructures}
\begin{document}

\begin{abstract}

Interface engineering in complex oxide heterostructures has developed into a flourishing field as various intriguing physical phenomena can be demonstrated which are otherwise absent in their constituent bulk compounds. Here we present La$_{0.7}$Sr$_{0.3}$MnO$_3$ (LSMO) / Sm$_{0.5}$Ca$_{0.5}$MnO$_3$ (SCMO) based heterostructures showcasing the dominance of antiferromagnetic interaction with increasing interfaces. In particular, we demonstrate that exchange bias can be tuned by increasing the number of interfaces; while, on the other hand, electronic phase separation can be mimicked by creating epitaxial multilayers of such robust charge ordered antiferromagnetic (CO-AF) and ferromagnetic (FM) manganites with increased AF nature, which otherwise would require intrinsically disordered mixed phase materials. The origin of these phenomena is discussed in terms of magnetic interactions between the interfacial layers of the LSMO/SCMO.
A theoretical model has been utilized to account for the experimentally observed magnetization curves in order to draw out the complex interplay between FM and AF spins at interfaces with the onset of charge ordering.

\end{abstract}

\section{Introduction}
The surfaces and interfaces of artificially layered complex oxide heterostructures frequently display novel properties that differ significantly from their bulk counterparts. Owing to some significant advances in the fabrication techniques of high quality oxide heterostructures \cite{hwang2012emergent, xia2020research, boileau2019textured}, and the structural compatibility of constituent complex oxides \cite{abdullah2018structural, liu2016strain}, the minute tuning of interfacial properties (which arise due to abrupt changes in electronic density, lowering of crystal symmetry, strain and/or defects at the interfaces) can be routinely achieved. These have led to a rapid surge in the investigation of a variety of fascinating phenomena like, interfacial magnetism \cite{yi2013tuning, lan2019interfacial, choi2021current}, magnetic frustration \cite{ding2016manganite, ding2013interfacial} and exchange bias \cite{chen2017hidden, sahoo2017interfacial, dong2009exchange}, orbital reconstruction \cite{zhou2020orbital, liang2017charge, deng2016modulating}, magnetoelectric coupling \cite{yi2013tuning, pesquera2020large} to name a few.

Among several complex oxide compounds, perovskite manganites have been a relevant class of materials for technological applications such as, advanced spintronic devices \cite{majumdar2013pulsed, volkov2012spintronics}, magnetic tunnel junctions (MTJs) \cite{jo2000very, mukhopadhyay2006inversion, sefrioui2010all}, magnetic memory devices \cite{ruotolo2007novel}, \textit{etc}. For example, La$_{0.7}$Sr$_{0.3}$MnO$_3$ (LSMO), being half--metallic ferromagnetic (FM) above room temperature (upto $\sim$ 350 K), has been the most frequently utilized FM layer in multilayer MTJs \cite{majumdar2013pulsed, mukhopadhyay2006inversion}. In the early stages, SrTiO$_3$, LaAlO$_3$, \textit{etc.} sandwiched as insulating barriers between LSMO layers in MTJs \cite{viret1997low, garcia2004temperature}; however, the results were not that encouraging, probably, due to more unstable and uncontrollable interfacial magnetization as compared to the more stable bulk magnetization.

Can one utilize all-manganite-based multilayers to study the interfacial magnetic behavior in a controlled manner? Infact to study the interfacial phenomena, during the last decade, researchers have started growing all manganite multilayers \cite{niebieskikwiat2006effect, mukhopadhyay2006giant, mukhopadhyay2008colossal}. This not only provides structural compatibility between each layers but also bringforth the phase compatibility [like presence of different kinds of antiferromagnetic (AF) phases or charge order (CO), insulating (I) phases] at the interfaces of the various manganites depending on the doping concentrations. For example,   Niebieskikwiat \textit{et. al.}\cite{niebieskikwiat2007nanoscale}, used antiferromagnetic (AF) Pr$_{0.7}$Ca$_{0.3}$MnO$_3$ (PCMO) compound as insulating layer in LSMO/PCMO heterostructures to study the interfacial magnetization behavior [by polarized neutron reflectometry (PNR) and exchange bias (\textit{EB})]. Performing depth profile of magnetization using PNR and conventional dc magnetometry, they showed that in nominally AF PCMO layer, maximum FM could be obtained when the PCMO layer thickness became comparable to the FM nanocluster size, which further hampered the saturation moment in LSMO interface layer. On the other hand, using LSMO/NSMO (NSMO = Nd$_{0.67}$Sr$_{0.33}$MnO$_3$) multilayers, where both the layers were FM, Mukhopadhyay \textit{et. al.} have shown giant enhancement of magnetoresistance \cite{mukhopadhyay2006giant}. Using LSMO/PCMO based FM/AF multilayers, they have also shown that these systems could mimic the transport
and magnetic characteristics of a spontaneously phase-separated manganites \cite{mukhopadhyay2008colossal}.
Ding \textit{et. al.} showed that competing magnetic interactions in bulk of FM LSMO and `G-type' AF SrMnO$_3$ led to magnetically frustrated interfacial spin configuration which further enhanced the $EB$ at LSMO/SMO interface \cite{ding2013interfacial}. Very recently, Chen \textit{et. al.} have shown, using PNR, the presence of a canted spin layer between the FM film and AF interfacial layer in single layer films of La$_{0.7}$Ca$_{0.3}$MnO$_3$ (LCMO) as well as LSMO, that led to negative as well as positive exchange bias upon different field cooling strengths \cite{chen2017hidden}. Similarly, Keunecke \textit{et. al.} have reported charge transfer induced mixed phase ferromagnetic behavior at the AF/AF interface composed of LaMnO$_3$/SrMnO$_3$, that mimicked high $T_{\rm{C}}$ ($T_{\rm{C}}$ = Curie temperature) FM behavior of optimally doped LSMO \cite{keunecke2020high}. In many of these reports, very sophisticated modern experimental techniques like PNR, annular bright field transmission electron microscopy (ABF-TEM), \textit{etc.}, have been used to directly probe the interfacial layers and their magnetization.
 
Besides, Miao \textit{et. al.} have shown experimentally that the thin films of chemical-doped (intrinsically disordered) La$_{1/3}$Pr$_{1/3}$Ca$_{1/3}$MnO$_3$ (LPCMO) exhibited pronounced large-length-scale phase separation, while perfectly ordered superlattice of LPCMO ([LaMnO$_3$/PrMnO$_3$/CaMnO$_3$]$_n$) (where all the individual layers are intrinsically AF and insulator) exhibited no signs of phase separation\cite{miao2020direct}. Their results indicate that chemical disorder / mixing is crucial for electronic phase separation. However, there seems to be very little investigation on interfacial magnetic behavior with the onset of charge ordering in heterostructures composed of robust charge ordered or CE-type antiferromagnetic manganites.

In this paper, we demonstrate the evolution of the interfacial antiferromagnetic behavior in all-manganite-based heterostructures composed of FM metallic compound La$_{0.7}$Sr$_{0.3}$MnO$_3$ (LSMO) and a robust charge ordered (CO) CE-type AF compound Sm$_{0.5}$Ca$_{0.5}$MnO$_3$ (SCMO) by investigating their magnetization as well as exchange bias loops at low temperatures. By fabricating multilayers with various number of interfaces, we describe the variation of interfacial magnetization orientation from the $M-T$ behavior and shape of $EB$ loops. In particular, the emergence of interfacial antiferromagnetic interaction and possible electronic phase separation due to dominance of charge ordering at low temperature with increasing multilayers is highlighted, which is recognized from atypical low-field $M-T$ as well as virgin loop $M-H$ behavior.
Our two-band model Hamiltonian calculations using spin-fermion Monte-Carlo method explains the atypical magnetization behavior, unveiling the role of interfacial magnetization of FM layer and charge ordering of AF layer.

\section{Experimental Details} 
A series of FM/AF heterostructures (named ML-1, ML-2 and ML-5) composed of $n$ repetitions of [LSMO($d_{FM}$)/SCMO($d_{AF}$)] bilayer ($d_{FM/AF}$ = thickness of FM / AF layer in nm) were grown from stoichiometric LSMO and SCMO bulk targets on TiO$_2$ terminated (001)-oriented SrTiO$_3$ (STO) substrates by pulsed laser deposition (PLD) technique using KrF excimer laser ($\lambda_{laser}$ = 248 nm). Further details of sample preparation are given in the \textit{Supporting Information}. The values of $d$ and $n$ were so chosen, such that, the total thickness of all the heterostructures remain fixed at $\sim$ 60 nm. Note that in all the heterostructures, LSMO was kept as the first deposited base-layer (as it exhibits better epitaxy with STO (001) substrate) and SCMO as the top termination layer. Thus, the number of interfaces were varied, keeping the total surface-to-volume ratio almost fixed (\textit{i.e.}, keeping the total thickness fixed to around 60 nm); the volume ratio of AF and FM layer was also kept fixed to $\sim$ 1.5. The descriptions of the three heterostructures so prepared have been mentioned in Supporting Table 2.

The total and individual layer thicknesses, crystal structures and epitaxy of the thin films have been analyzed by cross-sectional high resolution transmission electron microscopy (HRTEM) imaging in a TEM (FEI Tecnai G2 F30-ST) operated at 300 k$e$V.
The d.c. magnetization measurements of the films were performed using a commercial superconducting quantum interference device based vibrating sample magnetometer (SQUID-VSM) (MPMS-3 of M/s. Quantum Design, USA) in zero-field-cooled (ZFC) and field-cooled-warming (FCW) protocols. Details of various magnetization measurement protocols are given in \textit{Supporting Information}.

\section{Results and discussion}
\subsection{Characterizations}\label{sub1}
\begin{figure*}
\centering
\includegraphics[width=\columnwidth]{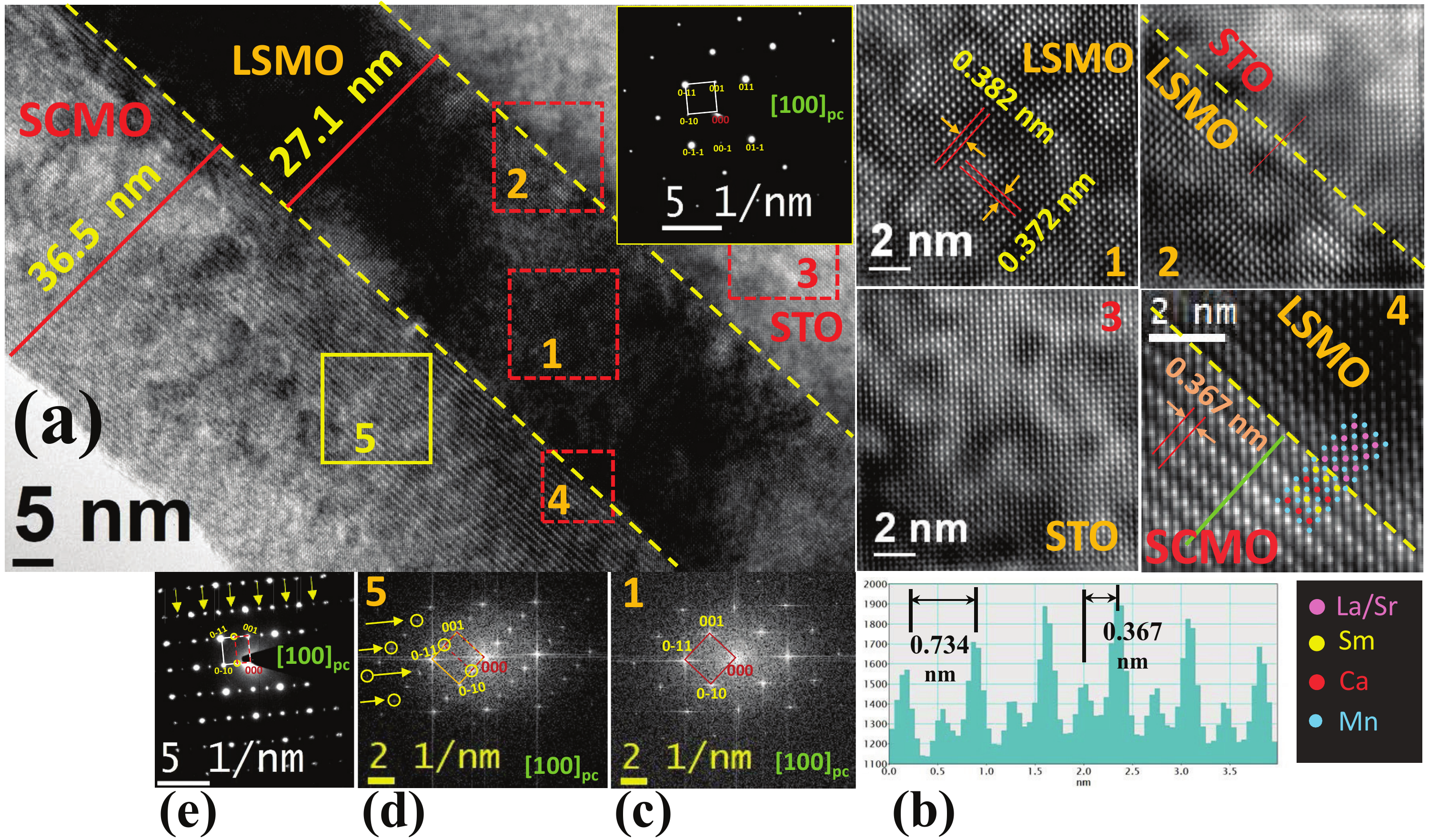}
\caption{\label{Fig_1_HRTEM_ML_1}(a) Cross-sectional HRTEM image of bilayer sample ML-1. [Inset: SAED pattern of STO substrate only.] (b) Magnified inverse Fourier filtered HRTEM images of selected areas marked in (a) [The line profile of SCMO layer along out-of-plane direction (marked by yellow dashed line in panel-4) has also been shown]. (c,d) Fourier transformed diffraction pattern of LSMO layer and SCMO layer respectively, obtained from the correponding marked regions in (a). (e)SAED patterns of substrate + film.}
\end{figure*}

\begin{figure*}
\centering
\includegraphics[width=0.97\columnwidth]{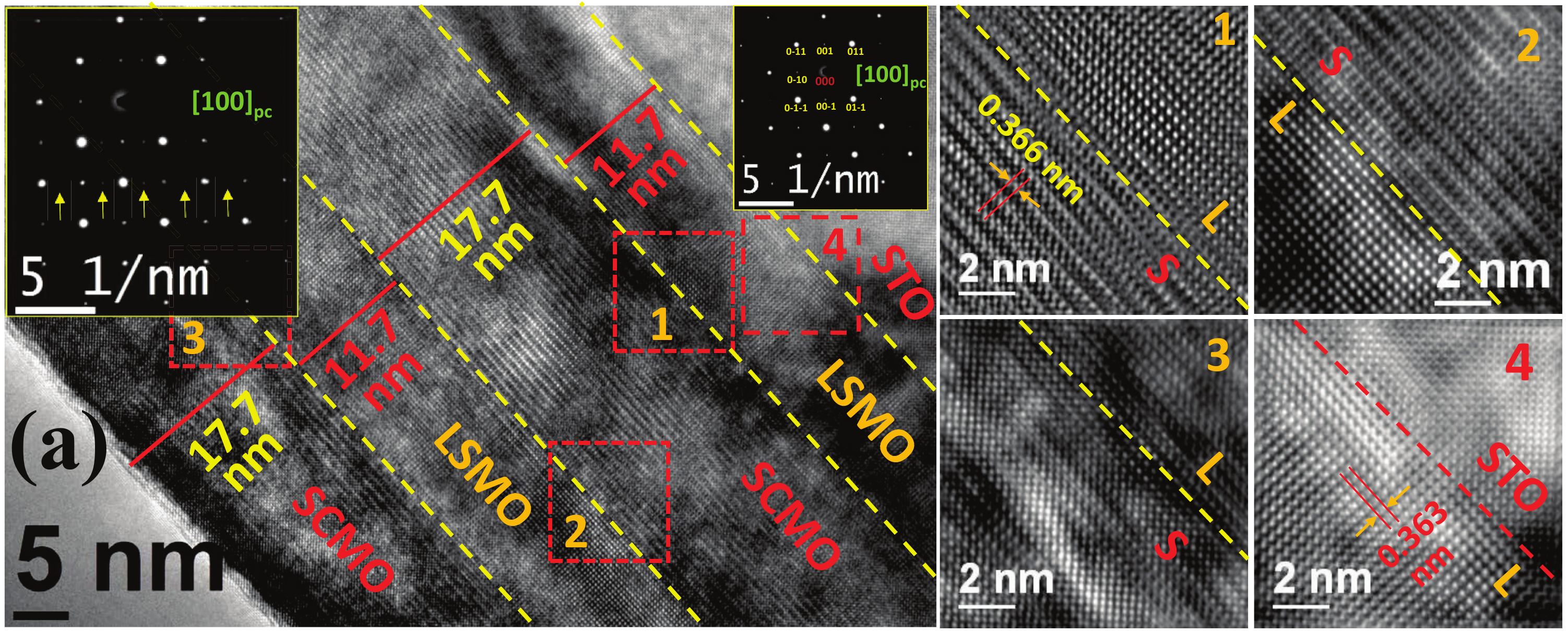}
\caption{\label{Fig_2_HRTEM_ML_2}(a) Cross-sectional HRTEM image of bilayer sample ML-2. [Inset: SAED patterns of the STO substrate only (top right) and substrate + film (top left).] On the right, magnified inverse Fourier filtered HRTEM images of selected areas marked in (a) are displayed (here `L' and `S' denote LSMO and SCMO layers respectively).}
\end{figure*}

\begin{figure*}
\centering
\includegraphics[width=0.97\columnwidth]{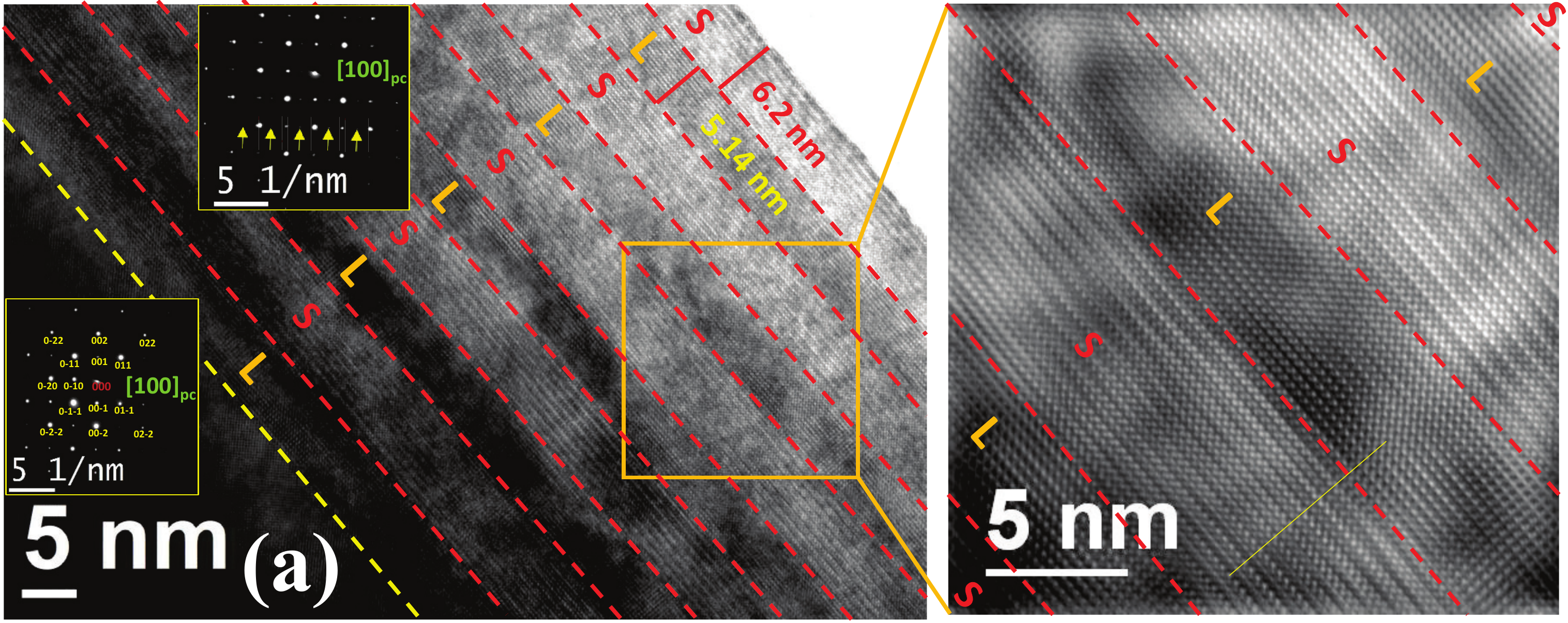}
\caption{\label{Fig_3_HRTEM_ML_5}(a) Cross-sectional HRTEM image of multilayer sample ML-5. Here `L' and `S' denote LSMO and SCMO layers respectively. [Inset: SAED patterns of the STO substrate only (middle left) and substrate + film (top).] Magnified inverse Fourier filtered HRTEM image of the selected area is shown separately on the right panel.}
\end{figure*}

The cross-sectional HRTEM images of the STO // [LSMO/SCMO]$_n$ heterostructures taken at room temperatures are shown in Fig. \ref{Fig_1_HRTEM_ML_1} (for ML-1), \ref{Fig_2_HRTEM_ML_2} (for ML-2) and \ref{Fig_3_HRTEM_ML_5} (for ML-5). These figures reveal excellent layer-by-layer growth and sharp interfaces between STO/LSMO and LSMO/SCMO layers in all the films and also confirms the absence of any amorphous layers. The thickness of each of the layers in each MLs are evident from respective figure panels (a). The experimentally obtained selected area electron diffraction (SAED) patterns are also shown in the insets of these figures (a). All the SAED patterns were captured along symmetric [100] zone axis (indexed in pseudocubic (\textit{pc}) notation) with respect to the substrate. Figure \ref{Fig_1_HRTEM_ML_1}(b) shows filtered images of the selected regions 1 to 4 marked in panel (a).

In addition, the SAED pattern of only the STO substrate along [100]$_{pc}$ zone axis is shown in the inset of Figure \ref{Fig_1_HRTEM_ML_1}(a) which reveals the cubic structure of the substrate. However, the SAED pattern of the total sample (film + substrate) taken along the same zone axis with respect to the substrate (panel \ref{Fig_1_HRTEM_ML_1}(e)) shows some extra spots (indicated by arrows). To further investigate their origin, fast Fourier transform (FFT) diffraction patterns of each layers are studied separately, in detail.

The FFT diffraction pattern of the LSMO layer (indexed along [100]$_{pc}$ in Fig. \ref{Fig_1_HRTEM_ML_1}(c), obtained from marked region-1 in panel (a)) reveals an epitaxial ``cube-on-cube" growth of LSMO layer on the STO substrate. This also means that the extra spots in SAED pattern in panel \ref{Fig_1_HRTEM_ML_1}(e) do not originate from the LSMO layer. Next, we focus on the FFT diffraction pattern of the SCMO layer shown in Fig. \ref{Fig_1_HRTEM_ML_1}(d) (obtained from region-5 in panel (a)) and observe some extra spots (indicated by arrows). Indexing the FFT pattern along [100]$_{pc}$ direction with respect to the substrate elucidates an orthorhombic ($Pnma$) growth structure of the SCMO layer.

Moreover, a contrasting feature is observed in the structure of SCMO layer as shown in filtered image in sub-panel \ref{Fig_1_HRTEM_ML_1}(b)-4. Along the out-of-plane direction, bright intensity atoms are separated at exactly 2 times the in-plane lattice parameter. This has been confirmed by the line profile along the out-of-plane (`$c$'-axis) lattice directions, as shown in figure panel \ref{Fig_1_HRTEM_ML_1}(b). However, further careful inspection of the profile and indicate that the spacing between low intensity and bright intensity spots along out-of-plane direction is equal to the in-plane lattice parameter. We thus model this as a very clean system with highly A-site ordered growth, like $\cdots$[--SmO--MnO$_2$--CaO--MnO$_2$--SmO--MnO$_2$--]$\cdots$ layered structure instead of $\cdots$[--Sm(Ca)O--MnO$_2$--Sm(Ca)O--MnO$_2$--]$\cdots$ layers \cite{akahoshi2003random, millange1998order, nakajima2002structures, tokura2006critical}, as shown by colored dots in sub-panel \ref{Fig_1_HRTEM_ML_1}(b)-4. We believe that atomic resolution $z$-contrast HAADF image would have given the best view of the atomic position.

Despite the differences in lattice parameters and crystal structures of STO ($a$ = 0.3905 nm, $Pm\bar{3}m$), bulk LSMO ($a$ = 0.5457 nm, $R\bar{3}c$) \cite{ding2013interfacial} and bulk SCMO ($a$ = 0.5415 nm, $b$ = 0.7548 nm, $c$ = 0.5360 nm, $Pnma$) \cite{rauwel2005stress}, HRTEM images and SAED patterns depict ``cube-on-cube" growth of the films even in the multilayers \ref{Fig_2_HRTEM_ML_2} and \ref{Fig_3_HRTEM_ML_5}. The structural properties of the SCMO layers persist even upto the top most layers in the multilayer films as confirmed from the SAED patterns of film + substrate, as shown in the insets of \ref{Fig_2_HRTEM_ML_2}(a) and \ref{Fig_3_HRTEM_ML_5}(a)

\subsection{Magnetic properties}\label{sub2}
\subsubsection*{$M - T$ variation}\label{sub_sub-2.1}
\begin{figure}
\centering
\includegraphics[width=\textwidth]{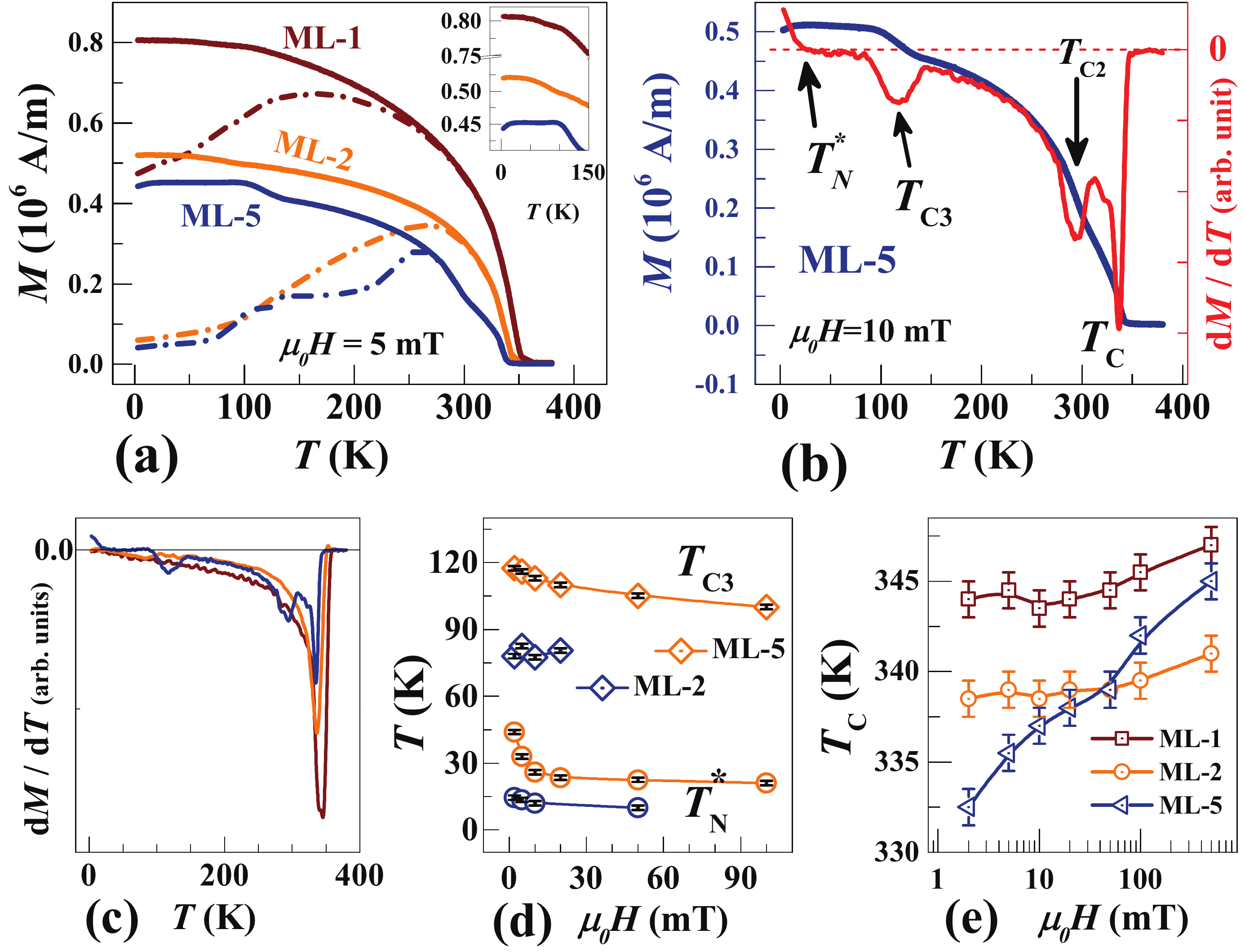}
\caption{\label{Fig_4_MT}(a)$M-T$ variation of the three samples measured at 5 mT (Broken curves: ZFC ; Solid curves: FCW) [Inset: Magnified view of the FCW curves at low $T$ region]. (b) $M-T$ (blue curve, left axis) and d$M$/d$T$ (red curve, right axis) variation of the FCW curve of ML-5 at 10 mT. The different transition temperatures are labelled. (c) d$M$/d$T$ variation of FCW curves of the three samples measured at 5 mT (corresponding to FCW $M-T$ curves in (a). Magnetic field variations of: (d) ($T_{\rm{N}}^*$) and $T_{\rm{C3}}$ for ML-2 (blue symbols) and ML-5 (orange symbols) extracted from corresponding FCW d$M$/d$T$ curves (solid lines are guide to the eye); and (e) FM transition (Curie) temperatures ($T_{\rm{C}}$) of all the samples.}
\end{figure}

The magnetization versus temperature ($M-T$) curves of the heterostructures (\textit{viz.}, ML-1, ML-2 and ML-5) measured in ZFC--FCW protocols (explained in \textit{Supporting Information}) under magnetic field of 5 mT are shown in Fig. \ref{Fig_4_MT}(a).  Firstly, we observe that the ferromagnetic transition (Curie) temperatures ($T_{\rm{C}}$) , obtained from the minima of the d$M$/d$T$ curves as shown in Fig. \ref{Fig_4_MT}(c), for all the films are close to the $T_{\rm{C}}$ of well-studied highly epitaxial ferromagnetic LSMO films \cite{ding2013interfacial, huijben2008critical}.
Secondly, a huge bifurcation between the ZFC and FCW curves below some certain temperatures can be observed. The temperature at which a ZFC curve starts to bifurcate from its corresponding FCW curve is commonly called the irreversibility temperature ($T_{irr}$). This kind of  bifurcation in $M-T$ below a Curie temperature is reminiscent of re-entrant spin glass (SG) behavior, where $T_{irr}$ denote the onset of the SG. It is further confirmed from the magnetic field variations of the $T_{irr}$, which follow de-Almeida Thouless (A-T) line \cite{deAlmeida1978stability, binder1986spin, deng2016modulating} (as described in the \textit{Supporting Information}, Fig. S 12(d)--(f)).
It thus remains to be established that the SG behavior occurs mainly because of the spin frustration at the interfaces. It is known that CE-type AF phase in SCMO is dominated by superexchange interaction between its Mn spins \cite{rao1998charge, tokura2014colossal, arulraj1998charge}, while FM in LSMO is due to double exchange interaction \cite{tokura2006critical, dorr2006ferromagnetic}. An increase in the number of interfaces ($N_i$) in the LSMO/SCMO heterostructure increases the competition between AF superexchange interactions involving Mn ions in SCMO and the FM double exchange interactions between Mn ions in LSMO, mainly at the interfaces. This leads to a large number of frustrated couplings between Mn spins at the interface of LSMO and SCMO. In fact, such a competition has been reported in many previous studies on manganite heterostructures \cite{ding2013interfacial, may2009enhanced}. At the FM/AF interface, the nearest neighbor Mn spins in the FM layer will always be influenced by an opposing pinning force from the Mn spins in AF layer. Moreover, the bifurcation also increases with increasing multilayers and $N_i$. Thus it can be claimed that this behavior originates mainly from the frustration of Mn spins at the interfaces of LSMO and SCMO layers.

However, there are some more prominent and distinct features in the magnetization of all the three films as evident from $M-T$ curves in Fig. \ref{Fig_4_MT}(a) and d$M$/d$T$ curves in  Fig. \ref{Fig_4_MT}(b)-(c).  These changes are characterized by different temperatures as shown, specifically for ML-5, in Fig. \ref{Fig_4_MT}(b). Clearly, the changes become more prominent as $d_{FM/AF}$ are decreased (or $N_i$ is increased). Moreover, these changes diminish upon application of higher magnetic fields and their variations are shown in Fig. \ref{Fig_4_MT}(d) as a function of magnetic field. A decrease in FCW magnetization at low fields and low temperature is observed (conventionally denoted by a N{\'e}el temperature $T_{\rm{N}}^*$). This $T_{\rm{N}}^*$ is most prominent in ML-5 (inset of Fig. \ref{Fig_4_MT}(a)) than in ML-2 and completely absent in ML-1. The appearance of $T_{\rm{N}}^*$ in $M-T$ might be attributed to the antiferromagnetic (AF) interaction between the Mn spins near the interfaces. Moreover, the cumulative FM ordering in LSMO dominates over the AF interaction at the single interface of the bilayer system ML-1, and thus leads to the absence of $T_{\rm{N}}^*$ for ML-1. Not only that, $T_{\rm{N}}^*$ further decreases with increasing cooling field in both ML-5 and ML-2 (shown in Fig. \ref{Fig_4_MT}(d)) and disappears above a certain cooling field. This behavior suggests that the Zeeman interaction in FM layers starts to predominate the AF interaction at higher cooling fields \cite{sahoo2018ultrathin}. We explore this further through M-H measurements, as discussed in the next subsection.

Another change in magnetization occurs at a comparatively higher temperature (denoted by $T_{\rm{C3}}$) $\sim$ 120 K for ML-5 and $\sim$ 80 K for ML-2. Finally, below $T_{\rm{C}}$, a change in magnetization occurs at $T_{\rm{C2}}$ ($\approx$ 294 K for ML-5) . In order to elucidate the origin of these changes and their relation to the interfacial magnetic interaction, we perform spin-fermion Monte Carlo calculations.

\subsubsection{Theoretical simulations}
We consider following two-band double exchange model Hamiltonian\cite{dagotto2001colossal} where Mn $e_g$ electrons (itinerant electrons) are Hund's coupled to the Mn $t_{2g}$ electrons (core spins) in a 3D cubic lattice:

\[H=\sum_{<ij>\sigma}^{\alpha \beta}(t_{\alpha \beta}^{ij}c_{i\alpha \sigma}^{\dagger}
c_{j\beta \sigma}+H.c)
-J_{H}\sum_{i}S_{i}.\sigma_{i}
-\lambda \sum_{i}Q_{i}.\tau_{i}
+J\sum_{<ij>}S_{i}.S_{j}\]
\[+\frac{K}{2}\sum_{i}{Q_{i}}^{2}-\mu \sum_{i}n_{i}\]

\noindent
Here, $c$ and $c^{\dagger}$ are the annihilation and creation operators for $e_{g}$ electrons, respectively. $<ij>$ implies nearest neighbor sites. $\alpha$, $\beta$ are summed over the two  Mn $e_{g}$ orbitals $d_{3z^{2}-r^{2}}$ and $d_{x^{2}-y^{2}}$ which are labeled as a and b, respectively. $t_{\alpha \beta}^{ij}$ implies the hopping amplitudes between $e_{g}$ orbitals on nearest neighbor sites
($t_{aa}^{x}=t_{bb}^{y}\equiv t,
t_{aa}^{y}=t_{bb}^{y}\equiv t/3,
t_{ab}^{x}=t_{ba}^{x}\equiv -t/\sqrt{3},
t_{ab}^{y}=t_{ba}^{y}\equiv t/\sqrt{3},
t_{aa}^{z}=t_{ab}^{z}=t_{ba}^{z}\equiv 0,
t_{bb}^{z}=4t/3$) where $x$, $y$ and $z$ are spatial directions. $J_{\rm H}$ is the Hund's coupling between $\sigma_i$ ($e_{g}$ spin) and ${\bf S}_{\rm i}$ ($t_{2g}$ spin). $\lambda$ denotes the coupling between the $e_{g}$ electrons to the Jahn-Teller phonons ${\bf Q}_{\rm i}$. We assume ${\bf S}_{\rm i}$ and ${\bf Q}_{\rm i}$ to be classical variables \cite{yunoki1998phase, dagotto1998ferromagnetic} and perform our calculations in $J_H \rightarrow \infty$ limit\cite{dagotto2001colossal, moreo2000giant}. In addition, we add an antiferromagnetic super-exchange interaction ($J$) between the classical spins. $K$ (stiffness of Jahn-Teller modes), $|{\bf S}_i|$, $t$ are set to be 1. We use a Monte Carlo technique based on travelling cluster approximation \cite{kumar2006travelling,pradhan2007distinct,pradhan2013electronic}.

In order to study our heterostructure system of LSMO (electron density $n_e$=0.7) and SCMO ($n_e$=0.5), the chemical potential (last term in the Hamiltonian) is tuned to set the desired electron density of the overall system along with the long range Coulomb (LRC) interaction between itinerant electrons \cite{yunoki2007electron,lin2008theory,pradhan2013electronic,pradhan2013interfacial}, which is essential to control the electron density of individual layers. LRC potentials restrain the amount of charge transfer among the layers. The strength of LRC interaction $\alpha$\cite{pradhan2013electronic,pradhan2013interfacial} is taken to be 0.5. We analyze the interfacial magnetic behavior of LSMO/SCMO heterostructures from one-to-one correspondence of the magnetization ($M-T$) process as seen in our experiments, with special emphasis to the ML-5 system.

First, we use $J=0.12$ and $\alpha=0$ (\textit{i.e.}, without using the LRC interaction) to reproduce the bulk LSMO-like and SCMO-like phases using $6\times6\times6$ calculations. We set $\lambda_L$=1.4 to simulate LSMO-like large bandwidth materials and obtained a FM metallic ground state for $n_e$=0.7 similar to experiments\cite{urushibara1995insulator, tokura1994giant, fujishiro1998phase, moritomo1998antiferromagnetic}.
For SCMO-like low bandwidth materials, we use $\lambda_S$=1.8 to attain the charge-ordered CE-type insulating phase at $n_e$=0.5 \cite{giri2014enhanced}.

Next, we design different type of LSMO/SCMO superlattices [$m$--LSMO/$p$--SCMO]$_{n}$ using the above parameter configurations. All our superlattice calculations are performed in a very small external magnetic field $h=0.005$. For this, an extra term $H_{mag}$ = $-h \sum_{i}S_i$ is added to the Hamiltonian. In superlattice structure [$m$--LSMO/$p$--SCMO]$_{n}$, $m$ ($p$) denotes the number of planes in LSMO (SCMO) layer (qualitatively equivalent to the individual layer thickness in experimental sample) while $n$ is the periodicity of the superlattice. In our calculation, applying travelling cluster approximation we access large system size ($4\times4\times60$) using $4\times4\times8$ travelling cluster.
As a result we have $(m+p) \times n$=60. We use three types of superlattices, namely $[27/33]_{1}$, $[13/17]_{2}$ and $[5/7]_{5}$ equivalent to our experimental samples ML-1, ML-2 and ML-5 respectively.
The z-component of the magnetization is calculated as $M_{z}=\frac{1}{N}\langle\sum_{i}S_{i}^{z}\rangle$, where $N$ is the number of sites. The angular brackets represent the thermal average over the Monte-Carlo generated equilibrium configurations. In addition, magnetization is averaged over ten different initial configurations of classical variables.

We set the LSMO/SCMO superlattice system using four values of electron-phonon coupling ($\lambda$). We represent the $\lambda$ for the interfacial LSMO (SCMO) plane by $\lambda_{LI}$ ($\lambda_{SI}$); whereas, the $\lambda$ used for rest of the LSMO (SCMO) layer is denoted as $\lambda_{L}$ ($\lambda_{S}$). Here we only consider one 2D plane of both LSMO and SCMO layers as the interface and denote the rest of the LSMO / SCMO layers barring the interfacial planes as bulk material.
As mentioned above, we use $\lambda_{L}$=1.4 and $\lambda_{S}$=1.8 to simulate LSMO and SCMO like materials. For the $MnO_{2}$ terminated LSMO layers the last plane of Mn ions of LSMO layers are coordinated by $La^{3+}$ and $Sr^{2+}$ in one side and $Sm^{3+}$ and $Ca^{2+}$ at the other side. Due to the smaller average ionic radius of $Sm^{3+}$ and $Ca^{2+}$ (in comparison to the average radius of $La^{3+}$ and $Sr^{2+}$ ions) the $\lambda$ value at the interfacial plane of LSMO layer ($\lambda_{LI}$) increases. But the $\lambda$ value at the interfacial plane of SCMO layer remains unaffected. For this reason we set $\lambda_{LI}$=1.6 and $\lambda_{SI}$=1.8.

\begin{figure*}
\includegraphics[width=\textwidth]{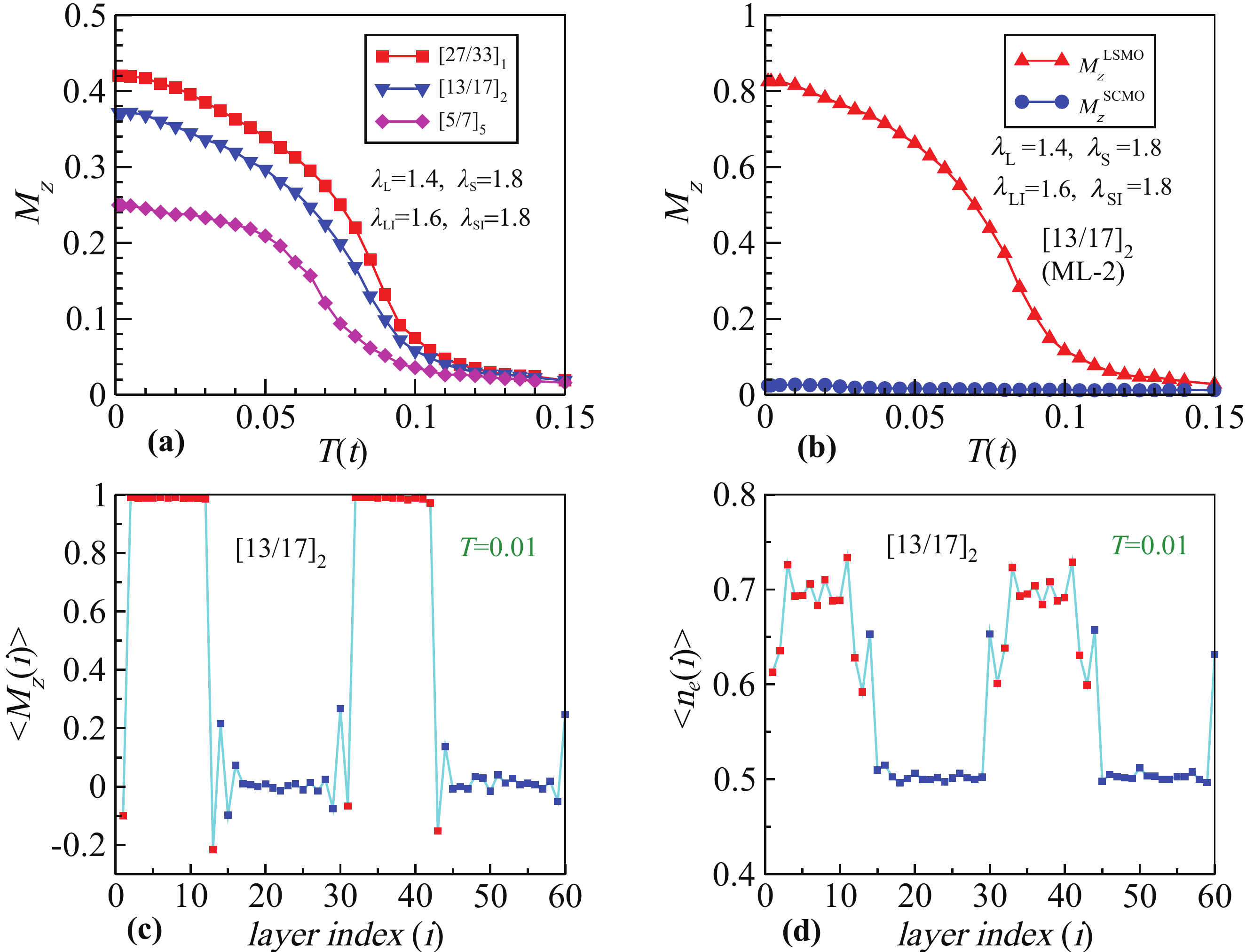}
\caption{
\label{1+1}
$\lambda$ configuration $\lambda_{\rm L}=1.4, \lambda_{\rm{LI}}=1.6,\lambda_{\rm S}=1.8, \lambda_{\rm{SI}}=1.8$: (a) Magnetization vs temperature profile for three different superlattices (ML-1, ML-2 and ML-5) are plotted. The saturation magnetization and $T_{\rm C}$ reduces with the decrease of LSMO layer width and increase of periodicity. (b) Magnetization vs temperature for the LSMO layers $(M_{z}^{\rm{LSMO}})$ and also for SCMO layers $(M_{z}^{\rm{SCMO}})$ in the $[13/17]_{2}$ SL system (\textit{i.e.}, ML-2) are displayed. The magnetization of the SCMO layers are almost negligible as compared to the LSMO layers. (c) Layer magnetization profile of the ML-2 system is presented at low temperature $T$=0.01. In the SCMO layers, the value of the magnetization is almost zero but the antiferromagnetic coupling is maintained at the interfaces. (d) Layer electron density of the ML-2 system is also presented at low temperature $T$=0.01. In the bulk LSMO and SCMO layers the required density maintains at 0.7 and 0.5 respectively but the density modified at the interfaces due to charge transfer at the interfaces.
}
\end{figure*}

The magnetization for three different superlattices using above mentioned parameters are shown in Fig.\ref{1+1}(a). The saturation magnetization at low temperature decreases with the increase of superlattice periodicity, as we go from $[27/33]_{1}$ (ML-1) to $[5/7]_{5}$ (ML-5). It is important to mention here that, like in our experimental samples, the total width of LSMO layers is more or same for three superlattices. The ferromagnetic transition temperature($T_{\rm{C}}$) also decreases due to the decrease of individual LSMO layer width as we increase the periodicity. These features of magnetization and $T_{\rm{C}}$ qualitatively agree well with the experiment results.

The total number of interfacial planes increases which brings down the saturation magnetization. In order to verify this we show the average magnetization of LSMO layers and SCMO layers separately in Fig.\ref{1+1}(b) for the $[13/17]_{2}$ (ML-2) system. This shows that the contribution from SCMO layer to the magnetization is minimal. So the dominant contribution to the magnetization of the systems (as shown in Fig.\ref{1+1}(a)) comes from the LSMO layers. But the average magnetization of LSMO layers (two LSMO layers for $[13/17]_{2}$ system) is below the saturation value. To understand this feature we plot the magnetization of each individual plane in Fig.\ref{1+1}(c). Magnetization of each plane in SCMO is very small as expected. The saturation magnetization is obtained only for LSMO bulk. 
However, the magnetization of interfacial LSMO planes decreases considerably. This decreases the magnetization of the LSMO layer as a whole as shown in Fig.\ref{1+1}(b). In case of $[13/17]_{2}$ superlattice the magnetization decreases for four interfacial LSMO planes. For $[27/33]_{1}$ ($[5/7]_{5}$) superlattice there are 2 (10) interfacial LSMO planes, where magnetization is affected. So with the increase of the periodicity the total number of interface increases that brings down the saturation magnetization as shown in Fig.\ref{1+1}(a). In addition to the magnetization of each plane we also plotted the average electron density $\langle n_e \rangle$ of each plane in Fig.\ref{1+1}(d). Due to the LRC interaction we find that the $\langle n_e \rangle$ of LSMO (SCMO) layers remains more or less at the desire density 0.7 (0.5) except for the interfacial planes. This indicate that the charge transfer is restricted to interfacial planes only in our calculations.

In the superlattices, particularly for $[5/7]_{5}$ superlattice (ML-5), we may have possibility of mixing of $La^{3+}$, $Sr^{2+}$, $Sm^{3+}$, and $Ca^{2+}$ ions  due to various experimental constraints like growth conditions, defects, etc., which are often beyond the control. In that scenario, interfacial planes of both the LSMO as well as the SCMO layers will be modified. For modelling such a case we set: 
$ \lambda_{L}=1.4,
\lambda_{LI}=1.5,
\lambda_{SI}=1.7,
\lambda_{S}=1.8 $.
%
%
\begin{figure*}
\includegraphics[width=\textwidth]{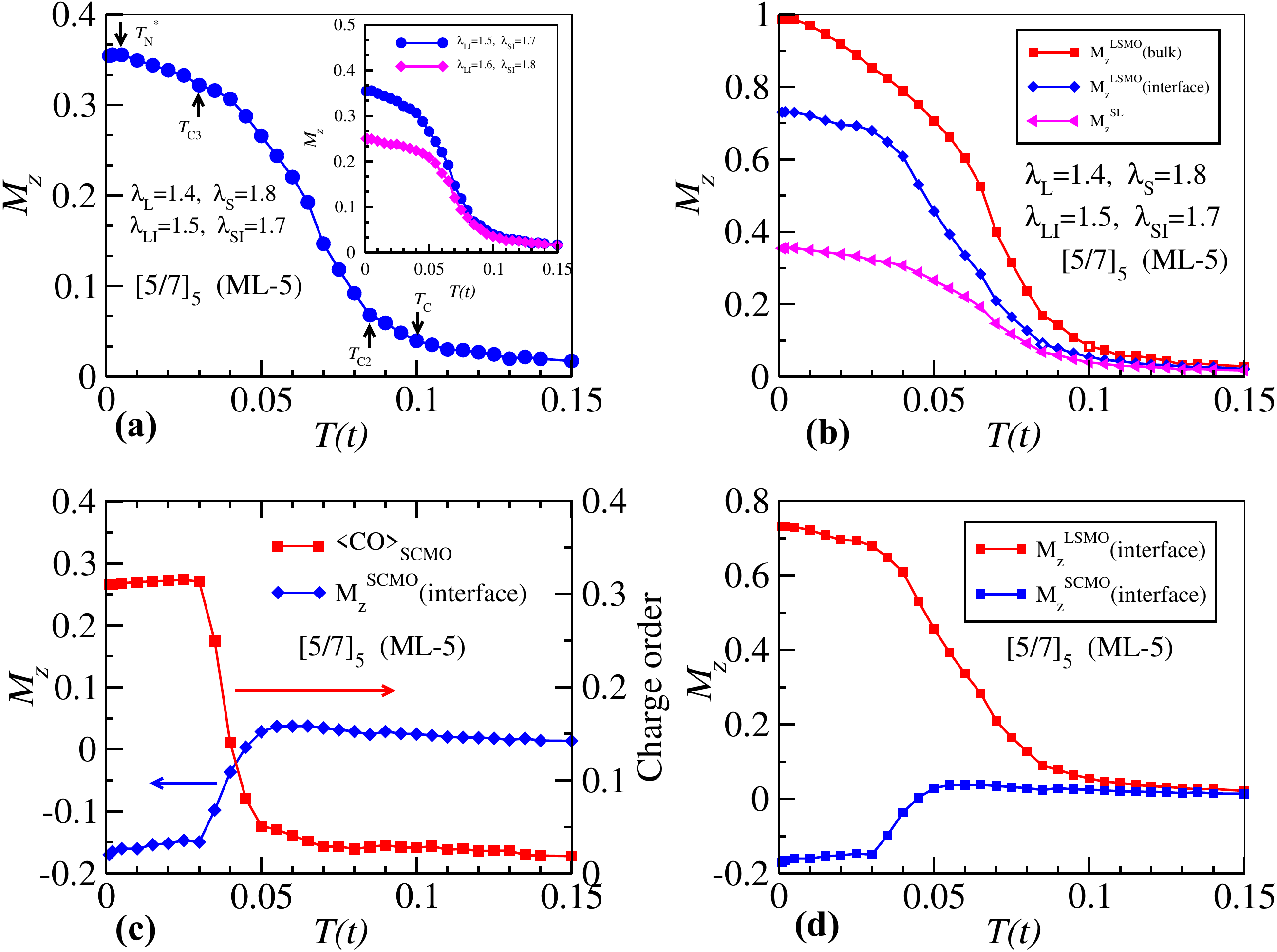}
\caption{
\label{5+5}
(a) Magnetization vs temperature of the $[5/7]_{5}$ SL system (ML-5) in the $\lambda$ configuration $\lambda_{\rm L}=1.4, \lambda_{\rm{LI}}=1.5, \lambda_{\rm S}=1.8, \lambda_{\rm{SI}}=1.7$ is plotted. Four steps appear in the magnetization curve at temperatures $T_{\rm{C}}$, $T_{\rm{C2}}$, $T_{\rm{C3}}$ and $T_{\rm{N}}^*$. In the inset, the magnetization of this $\lambda$ configuration is compared with the interface configuration $\lambda_{\rm{LI}}=1.6, \lambda_{\rm{SI}}=1.8$. (b) Comparision of the total ML-5 system's magnetization ($M_{z}^{total}$) with the interface LSMO layer magnetization [$M_{z}^{\rm{LSMO}}$(interface)] and the bulk LSMO layer magnetization (except the interfacial planes) [$M_{z}^{\rm{LSMO}}(bulk)$] is demonstrated. (c) The SCMO interface layer magnetization ($M_{z}^{\rm{SCMO}}$) is compared with its charge order ($\langle CO\rangle_{\rm{SCMO}}$). (d) Interface LSMO layer magnetization [$M_{z}^{\rm{LSMO}}$(interface)] and the interface SCMO layer magnetization [$M_{z}^{\rm{SCMO}}$(interface)] of the ML-5 system is presented. At the temperature point where SCMO interface layer magnetization start to increase, a slow-growth of LSMO interface layer magnetization is observed.
}
\end{figure*}
The magnetization of the $[5/7]_{5}$ superlattice for this scenario is plotted in Fig.\ref{5+5}(a). The scenario we discussed earlier in Fig.\ref{1+1}(a) is also plotted in the inset for comparison. The magnetization at low temperature increases as compared to the earlier case (see the inset).

We further analyze the magnetization of $[5/7]_{5}$ superlattice (shown in Fig.\ref{5+5}(a)) in more details. It is clear that the magnetization curve goes through a four-step-process. The first step appears at the ferromagnetic transition temperature $T_{\rm C}\sim 0.1$. Then the second step appears at $T_{\rm{C}2}\sim 0.085$. Finally the third step emerges at $T_{\rm{C}3}\sim 0.03$. At very low temperature, around $T\sim 0.01$ the magnetization takes a slight downward turn, similar to $T_{\rm{N}}^{*}$ in experimental $M-T$ curves. The four-step magnetization curve observed in experiment is quite well captured in our model Hamiltonian calculations.

In order to interpret details of the steps in magnetization data, first we have evaluated the magnetization of bulk LSMO layers (except the interfacial planes), interfacial LSMO planes and the total superlattice system which are as shown in Fig.\ref{5+5}(b).
The bulk LSMO layer transition temperature corresponds well with the transition  temperature of the superlattice at $T_{\rm C}\sim0.1$ which is reflected as the step in the magnetization curve in Fig.\ref{5+5}(a). The transition temperature of the interfacial LSMO planes is a bit lower than the bulk LSMO layer and lead to the second step in the magnetization curve (around $T_{\rm{C}2}\sim 0.085$). The magnetization of interfacial LSMO planes dips around $T_{\rm{C}3}\sim 0.03$ which contributes to the formation of the third step in magnetization in Fig.\ref{5+5}(a).

We next analyze the reason behind the slow-change in the magnetization just above $T_{\rm{C}3}$ in the interfacial LSMO planes. For this, we plot the magnetization of SCMO interfacial planes along with the charge ordering in the SCMO planes in Fig.\ref{5+5}(c). The onset of charge ordering in SCMO layer and alignment of spins in interfacial plane coincide with each other. Basically the magnetization in the interfacial SCMO layers increases simultaneously with the charge ordering and also gets saturated with the saturation of the charge ordering. It has to be emphasized here that, due to antiferromagnetic interaction at the interface the magnetization of interfacial SCMO layer orients opposite to the magnetization of the interfacial LSMO plane at low temperature as shown in Fig.\ref{5+5}(d). The onset of magnetization in interfacial SCMO plane due to the antiferromagnetic coupling with interfacial LSMO plane slows down the magnetization alignment of interfacial LSMO plane with bulk LSMO at around $T=0.04$. Once the magnetization of SCMO layers gets saturated the magnetization of interfacial LSMO start to increase at $T = 0.03$. At very low temperature ($T\approx0.01$), the downward turn of $M-T$ is most probably due to the small drop in the CO and magnetization of SCMO layers which consequently enhances the magnetization of the LSMO interface layers slightly. Overall, these calculations qualitatively explain our experimental $M-T$ curves taking into account the role of bulk and interfacial magnetization of LSMO and SCMO layers and CO of SCMO layer.

\subsubsection*{$M - H$ variation}\label{sub_sub-2.2}
\begin{figure}
\centering
\includegraphics[width=\textwidth]{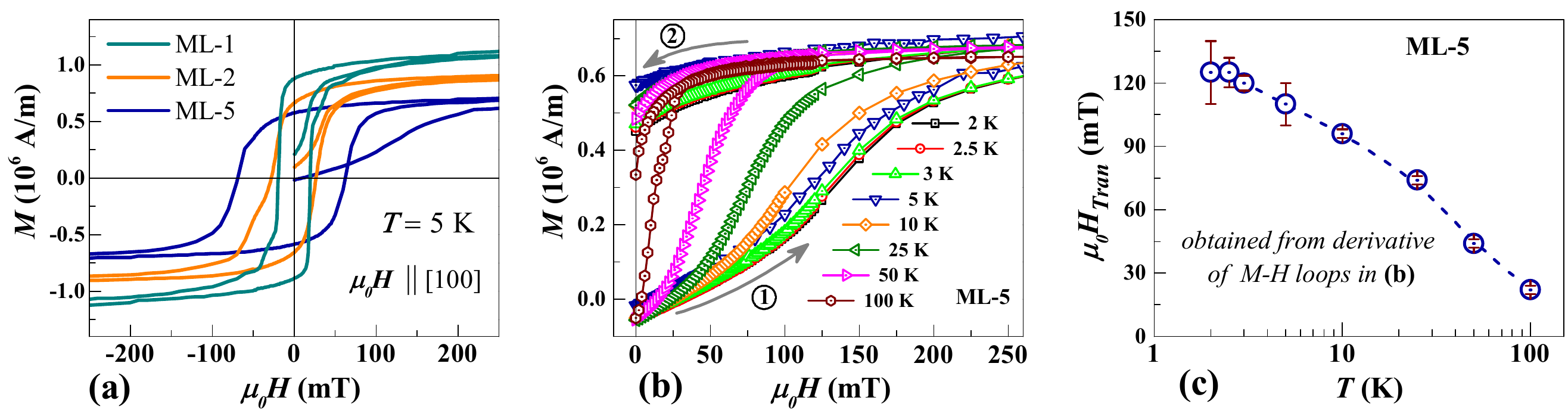}
\caption{\label{Fig_6_MH}(a) Low field (upto 0.2 T) magnetic hysteresis ($M-H$) loops of ML-1, ML-2 and ML-5 measured at 5 K along in-plane [100] direction. Note the behaviour of the virgin curve, which tends to go more out of the `loop' as the multilayer is increased (most prominent in ML-5). (b) 2-quadrant $M-H$ loops of ML-5 at various temperatures showing the variation of change in virgin curve behaviour. (c) Temperature variation of $\mu_0H_{Tran}$ (obtained from maxima of the slope of virgin curves).}
\end{figure}

Now in order to elucidate the competing AF interactions at the low $T$ regime, we performed ZFC hysteresis loop measurements at $T < T_{\rm N}^{*}$.
Figure \ref{Fig_6_MH}(a) shows the ZFC $M - H$ loop behaviour in the low field region for the three heterostructures measured along in-plane symmetric [100]$_{pc}$ direction at $T$ = 5 K. With increase of multilayers (\textit{i.e.}, $N_i$), three distinctive features can be observed: (i) coercivity increases; (ii) saturation magnetization ($M_{Sat}$) decreases; and (iii) the virgin curve goes more outside of the subsequent hysteresis loops.

Since the cumulative FM layer thickness is almost the same in all three samples, the increase in coercivity in this case is more likely an interfacial effect, which can be attributed to the increase of pinning with increasing number of interfaces  \cite{anyfantis2022growth}. Consequently, interfacial effect also controls the $M_{Sat}$ and the initial value of $M$ [\textit{i.e.} $M$($\mu_0H$ = 0.0 T)].
As $N_i$ increases, the number of interfacial spins not contributing to the total magnetization increases, which leads to such effects\cite{zhang2022temperature}. This might be due to a strong AF coupling at the interfaces, especially at low temperatures. Finally, the behaviour of virgin curve going outside of the $M-H$ loop (clearly visible for ML-5) also hints towards the increasing AF interaction. This virgin curve behaviour is also reminiscent of the magnetic field-induced metamagnetic transition, which is quite common in electronically phase separated manganites \cite{shao2016emerging}. Though a field-induced metamagnetic transition is generally quite abrupt in nature (first order transition) \cite{banik2018huge}, here in our case, in sample ML-5, the transition is quite smooth. Thus in order to find the exact transition field value $\mu_0H_{Tran}$, we have taken the field derivative [d$M$/d($\mu_0H$)] of the virgin curve, where the maxima denotes the transition \cite{curiale2007magnetism}. With increasing temperature, this maxima is found to decrease to a lower field, \textit{i.e.}, the value of $\mu_0H_{Tran}$ decreases (as in Fig. \ref{Fig_6_MH}(c)). These results confirm that the competition between the charge ordered AF phase and FM phase (as in intrinsically phase separated manganites) increases with increasing interfaces. This also suggests that one can induce the phase separation artificially by creating multilayers.

Now after speculating the above facts, can we experimentally comfirm the AF interaction among the interfacial spins across the interfaces (\textit{i.e.}, interfacial exchange interaction)?
\subsection{Exchange Bias}\label{sub3}
\begin{figure}
\centering
\includegraphics[width=\textwidth]{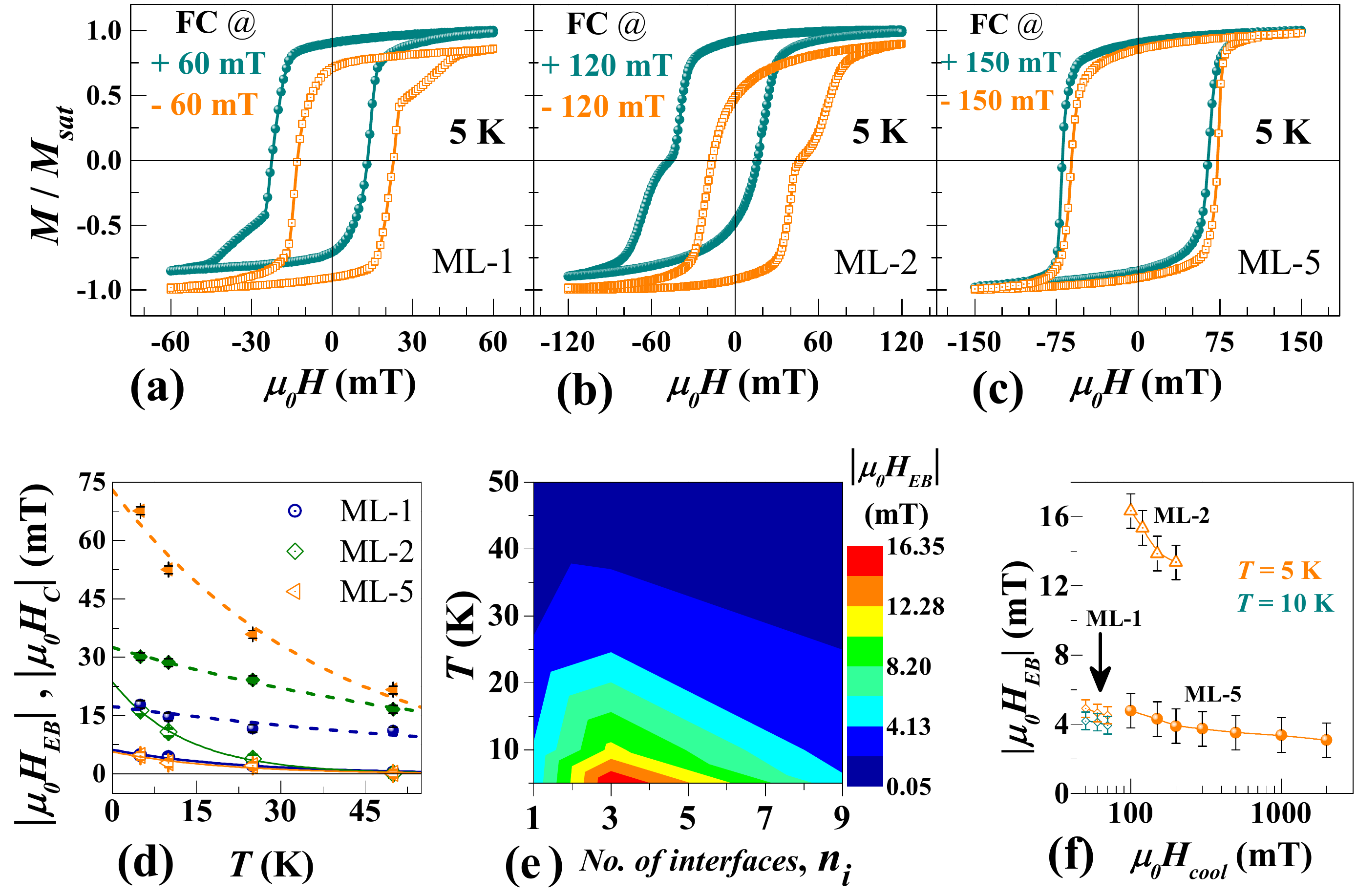}
\caption{\label{Fig_7_EB}Exchange bias ($EB$) hysteresis loops measured (along in-plane [100] direction) under positive (blue) and negative (orange) cooling fields at 5 K for: (a) ML-1, (b) ML-2, (c) ML-5 (Corresponding values of cooling fields are mentioned in the respective figure). (d) Temperature variations of $EB$ field (open symbols, solid curves) and coercive field (Solid symbols, dotted curves) of the three films. (The curved lines in (d) and (e) are fits to Eqs. \ref{Eqn_T_var}.) (e) Intensity plot showing variation of $EB$ field as a function of temperature and no. of interfaces ($N_i$). (f) Variation of $EB$ fields as a function of cooling fields measured (along in-plane [100] direction) for the three MLs.}
\end{figure}
To get better insight into the competing interfacial magnetic interactions in the LSMO/SCMO heterostructures we study the evolution of $EB$ effect. We follow the customary (magnetic) field-cooling techniques to generate the $EB$ hysteresis loops. We field-cool (FC) all the samples from $T_{FC}$ = 320 K. The $EB$ loops are measured at $T$ = 5 K after positive and negative field cooling and are shown in Fig. \ref{Fig_7_EB}(a)-(c).
Similar to ZFC $M - H$ loops, the coercive field ($\mu_0H_C$) in FC $EB$ loops also increases with increasing $N_i$. Apart form that, two other prominent features can be observed upon comparison: (i) the amount of exchange bias field ($\mu_0H_{EB}$) increases from ML-1 to ML-2 but diminishes in ML-5; (ii) the shapes of the $EB$ loops are distinctively different in each of these samples. To gain further insight into the evolution of $EB$ effect in each samples, temperature variation (at fixed cooling field, $\mu_0H_{cool}$) and cooling field variation (at fixed temperature) of $EB$ loops were performed.
 
Indeed, the temperature dependencies of $\mu_0H_{EB}$ and $\mu_0H_C$ of all the samples follow exponential decay, that can be fitted by the phenomenological formula given by \cite{moutis2001exchange}:
\begin{subequations}\label{Eqn_T_var}
\begin{equation}
\mu_0H_{EB}(T) = \mu_0H_{EB}^0 ~ \exp \left( -\frac{T}{T_1}\right)
\label{Eqn_T_var:H_EB}
\end{equation}
\begin{equation}
\mu_0H_C(T) = \mu_0H_C^0 ~ \exp \left( -\frac{T}{T_2}\right)
\label{Eqn_T_var:H_C}
\end{equation}
\end{subequations}
where $\mu_0H_{EB}^0$ and $\mu_0H_C^0$ are the extrapolations of $\mu_0H_{EB}$ and $\mu_0H_C$ to the absolute zero temperature; $T_1$ and $T_2$ are constants. This kind of exponential decay behaviours of $\mu_0H_{EB}$ and $\mu_0H_C$, shown in Fig. \ref{Fig_7_EB}(d), give further support to the fact that the $EB$ in the LSMO/SCMO bilayer can be attributed to the interface spin competition between the AF super-exchange and the FM double-exchange interactions \cite{ding2013interfacial, moutis2001exchange}.

On the other hand, according to Meiklejohn and Bean's model \cite{meiklejohn1956new}, magnitude of exchange bias in a standard FM/AF heterostructure can be written (under simplified conditions) \cite{nogues1999exchange} as Eq. \ref{Eqn_EB_vs_d_FM}:
\begin{equation}
\mu_0H_{EB} = \frac{E_{ex}}{M_{sat}d_{FM}}
\label{Eqn_EB_vs_d_FM}
\end{equation}
where $E_{ex}$ is the exchange energy density per unit volume across the FM/AF interface, $d_{FM}$ and $M_{sat}$ are the cumulative thickness and the saturation magnetization per unit volume of the FM layer, respectively. This implies that exchange bias field is inversely proportional to FM layer's thickness and magnetization. The calculated values of $E_{ex}$ for each samples (given in Table \ref{Tab_2}) are in good agreement with some reported results on this kind of oxide heterostructure systems \cite{murthy2017interface}.
\begin{table}[t]
\caption{Various parameters obtained from magnetization and exchange bias measurements:}
\label{Tab_2}
\begin{tabular}{cccccc}
\hline
Sample & \begin{tabular}[c]{@{}c@{}}$M_{Sat}$ ($\times$10$^6$)\\ at 5 K\end{tabular} & \begin{tabular}[c]{@{}c@{}}$\mu_0 H_{EB}^0$\\ (mT)\end{tabular} & \begin{tabular}[c]{@{}c@{}}$\mu_0 H_{C}^0$\\ (mT)\end{tabular} & \begin{tabular}[c]{@{}c@{}}$d_{FM}$\\ (nm)\end{tabular} & \begin{tabular}[c]{@{}c@{}}$E_{ex}$ at 5 K\\ (mJ.m$^{-2}$)\end{tabular} \\ \hline
ML-1 & 1.39 A/m & 6.2$\pm$0.5 & 17$\pm$1 & 27.1 & 0.180 \\
ML-2 & 0.97 A/m & 24$\pm$1 & 32.5$\pm$0.5 & 23.4 & 0.370 \\
ML-5 & 0.81 A/m & 5.7$\pm$0.8 & 73$\pm$3 & 25.5 & 0.096 \\ \hline
\end{tabular}
\end{table}

In bilayer system (ML-1), the main contribution to magnetization ($M$) comes from LSMO layer and the SCMO layer tries to pin the LSMO spins at the interface, a conventional $EB$ scenario. The d$M$/d$\mu_0H$ curves of the $EB$ plots (supporting Fig. S 13, lower panels) indicate that after reversal of the majority FM spins in the LSMO layer, the interfacial spins of LSMO reverse slowly with further increasing the field in opposite direction.
Next, in the four-layer system (ML-2), the main  contribution to $M$ still comes from bulk LSMO layer. But although $d_{AF}$ decreases, the CO-AF state of SCMO layer stabilizes; thus with increasing $N_i$ the SCMO layers pin the LSMO interfacial spins more strongly showing increase in coercivity. Consequently, a sharper transition/reversal of the interfacial LSMO spins occur at higher fields (as compared to that in ML-1). It is expected that with further increasing multilayers, $EB$ field should increase.
However, interesting phenomena occur when $d_{FM/AF}$ are further reduced and no. of interfaces ($N_i$) are further increased (say, in ML-5). Although, the FM moments of the bulk LSMO layers contribute to the $M$, the volume fraction of spins of the LSMO interfacial layers and that of bulk LSMO layers become comparable. But since $d_{FM}$ is quite small, the SCMO layers can now pin the interface spins of LSMO layers strongly and thus the AF dominance increases.

Although the AF nature of the system increased in ML-5, leading to tendency of phase separation, the non-monotonous behaviour in $EB$ is observed.

Also, variation of the $EB$ field for different cooling fields ($\mu_0H_{cool}$) are investigated where $\mu_0H_{cool}$ were chosen below and above $\mu_0H_{trans}$, specifically for ML-5. Such cooling field dependence for all the samples are shown in Fig. \ref{Fig_7_EB}(f). From this figure, it becomes clear that $H_{EB}$ decreases with increasing the cooling fields for all the samples. This can happen only when the interfacial spin of LSMO become AF aligned with that of the FM spin of the LSMO. If the number of FM layer become comparable with the AF aligned interfacial layer of LSMO, the EB decreases. Overall, the EB loop shape and the variation with the cooling field and $N_I$ signifies experimentally the AF interfacial interaction at the LSMO/SCMO interfaces.
%

\section{Conclusion}
To conclude, we have prepared a sets of multilayer heterostructures comprising of LSMO/SCMO by repeating this bilayer structure while keeping total thickness of the whole films fixed at $\sim$ 60 nm, thereby increasing the number of interfaces ($N_i$). Cross-sectional HRTEM of the three samples revealed excellent epitaxy with sharp interfaces even in the highest multilayer film (ML-5). Magnetization measurements revealed increase of interfacial competing interactions with increase of the number of multilayers as depicted by the multiple steps in the $M-T$ curves. Theoretical simulations were carried out to explain the origin of the atypical magnetization behavior, which revealed the dominant role of interfacial magentization at the LSMO/SCMO interfaces and the charge ordering of SCMO layers with the increase in multilayers. These theoretical results matched qualitatively well with the observed experimental results. Moreover, the increase in antiferromagnetic (AF) interaction in ML-5 at low temperatures was confiremed further from $M-H$ hysteresis loop measurements. Finally, exchange bias ($EB$) measurements revealed the possible magnetic interactions at the LSMO/SCMO interfaces. All-in-all, tendency of AF interaction increased with increasing multilayers. In this manner, one can tune the interfacial magnetic interaction by harnessing the CO-AF states in highly ordered epitaxial manganite superlattices.
\begin{acknowledgement}
The work was supported by Department of Atomic Energy (DAE), Govt. of India.
\end{acknowledgement}
\begin{suppinfo}\label{SI}
Details of sample preparation; procedure for eliminating remnant magnetic field in SQUID magnetometer for low field measurements, ZFC and FCW protocols; d$M$/d$T$ curves at various fields for all MLs; A-T lines for all MLs; $EB$-loop derivatives; Model Hamiltonian and methods. 
\end{suppinfo}
\section*{Supporting Information}
\section{Details of sample preparation}

Bulk polycrystalline targets of La$_{0.7}$Sr$_{0.3}$MnO$_3$ (LSMO) and Sm$_{0.5}$Ca$_{0.5}$MnO$_3$ (SCMO) compounds were synthesized using conventional sol-gel reaction method.

Prior to film deposition, the STO (001) substrates were treated in buffered NH$_{4}$F-HF (BHF) solution (of pH $\approx$4.5) followed by cleaning with DI water and dry N$_2$ blow and subsequent annealing at 950 $\degree$C for 2 hours in ambient O$_2$ environment to obtain TiO$_{2}$ terminated surface for better epitaxy of the films \cite{kawasaki1994atomic, biswas2011universal}.

For all three samples, the LSMO and SCMO layers were deposited at substrate temperature of 800 $\degree$C with O$_2$ partial pressure of about 0.6 mbar. During deposition, the target-to-substrate distance was 4 cm, the laser energy density at the target surface was 2 J.cm$^{-2}$ and the laser pulse repetition rate was kept at 1 Hz. After completion of the deposition, the chamber was filled up with O$_2$ up to 1000 mbar pressure and then the samples were cooled down at a rate of 10 $\degree$C/min. Schematics of the three heterostructures are given below.
\begin{figure}[H]
\renewcommand{\figurename}{Figure S}
\centering
\includegraphics[width=0.75\textwidth]{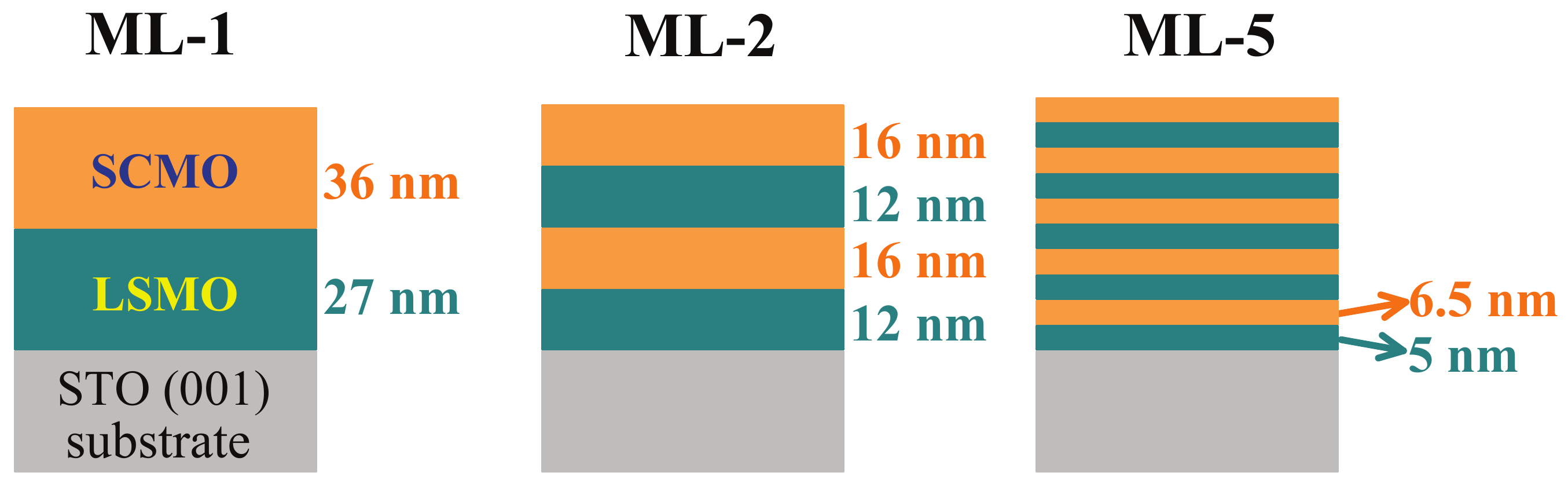}
\caption{\label{Fig_Supp_MLs}Schematics of the three heterostructures prepared.}
\end{figure}
\begin{table}[h]
\caption{Description of the prepared heterostructure. (Thickness of the films were obtained from cross-sectional HRTEM images.)}
\label{Tab_1}
\resizebox{\textwidth}{!}{%
\begin{tabular}{cccccc}
\hline
Film stack &
  \begin{tabular}[c]{@{}c@{}}Thickness of\\ whole film, $d$\end{tabular} &
  \begin{tabular}[c]{@{}c@{}}No. of repetitions\\ of the bilayer, $n$\end{tabular} &
  \begin{tabular}[c]{@{}c@{}}Total no. of\\ layers, $N_L$\end{tabular} &
  \begin{tabular}[c]{@{}c@{}}No. of FM/AF\\ interfaces, $N_i$\end{tabular} &
  \begin{tabular}[c]{@{}c@{}}Code\\ name\end{tabular} \\ \hline
STO // LSMO/SCMO       & 63.6 nm & 1 & 2  & 1 & ML-1 \\
STO // [LSMO/SCMO]$_2$ & 60.2 nm  & 2 & 4  & 3 & ML-2 \\
STO // [LSMO/SCMO]$_5$ & 57 nm  & 5 & 10 & 9 & ML-5\\
\hline
\end{tabular}%
}
\end{table}

\section{Magnetization Measurements}
During the in-plane configuration a flat surface quartz paddle was used, while for the out-of-plane configuration clear drinking straws (with background signal contribution $\sim$ 10$^{-8}$ emu, specifically provided by the manufacturer) were used. Temperature variations of magnetization were measured in two different protocols:

(i) zero-field-cooled (ZFC) protocol: Samples were cooled from 380 K down to low temperature (3 K) in the absence of external magnetic field (without acquiring the data). Then, a desirable constant magnetic field was applied at $T$ = 3 K and the magnetization data were recorded while warming from 3 K to 380 K;

(ii) field-cooled-warming (FCW) protocol: Consecutively after the ZFC measurement, samples were again cooled from 380 K down to 3 K, without turning ``off" the magnetic field. Thereafter, magnetization data were recorded while warming from 3 K upto 380 K, under same magnetic field.

\begin{figure}[H]
\renewcommand{\figurename}{Figure S}
\centering
\includegraphics[width=0.65\columnwidth]{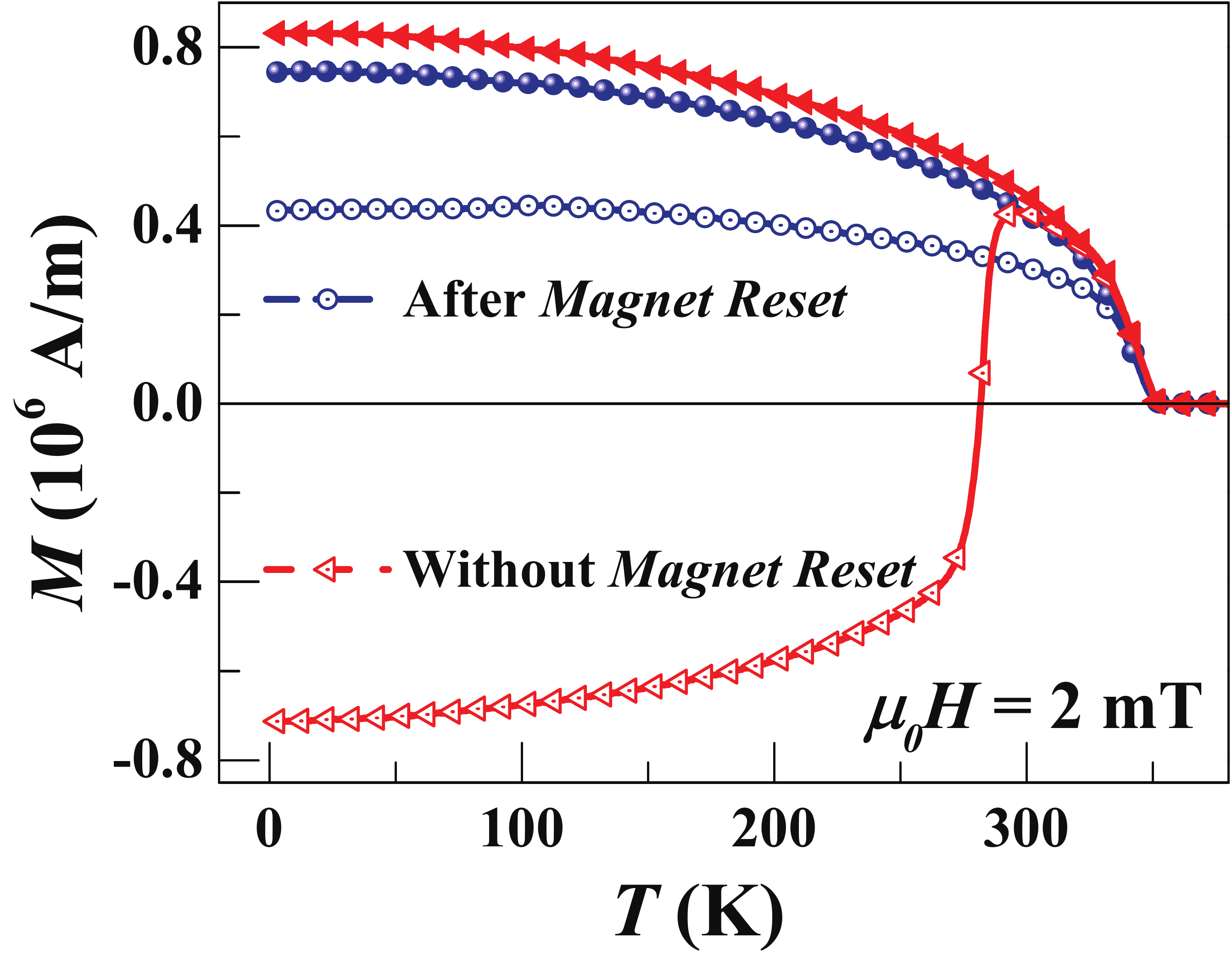}
\caption{\label{Fig_Supp_Magnet_Reset} Measured $M-T$ data for ML-1 at low applied magnetic field of 2 mT, before and after quenching the magnet. The effect of remnant field can be clearly seen for such soft FM samples.}
\end{figure}

To remove any artifact of remnant magnetic field (as shown in Fig. S \ref{Fig_Supp_Magnet_Reset} for ML-1 sample) arising from trapped current in Superconducting magnet of the VSM system, magnetic field oscillations followed by quenching of the magnet was performed when required.

\subsection{d$M$/d$T$ curves}
\begin{figure}[H]
\renewcommand{\figurename}{Figure S}
\centering
\includegraphics[width=\columnwidth]{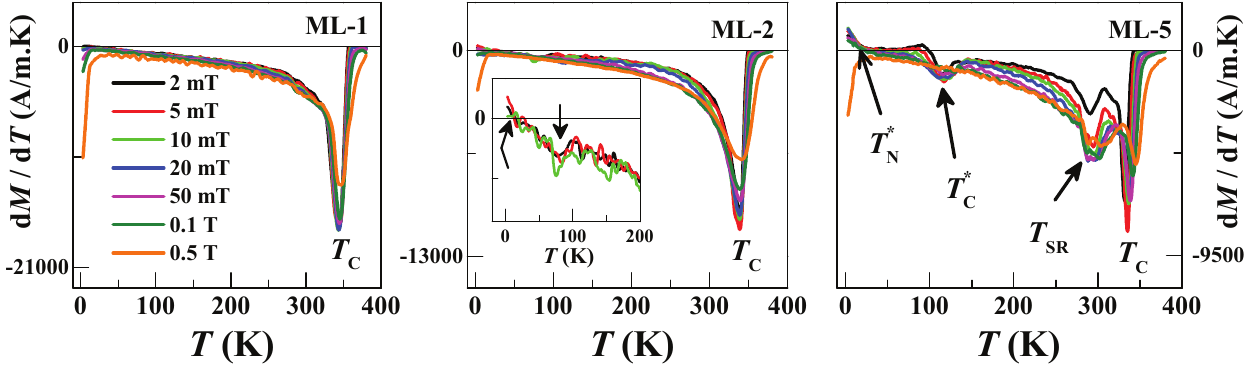}
\caption{\label{Fig_Supp_dM-dT} d$M$/d$T$ variation of ML-1, ML-2 and ML-5 at various fields. The transition temperatures are marked in respective figures }
\end{figure}
\subsubsection{Almeida-Thouless line}
\begin{figure}[H]
\renewcommand{\figurename}{Figure S}
\centering
\includegraphics[width=\columnwidth]{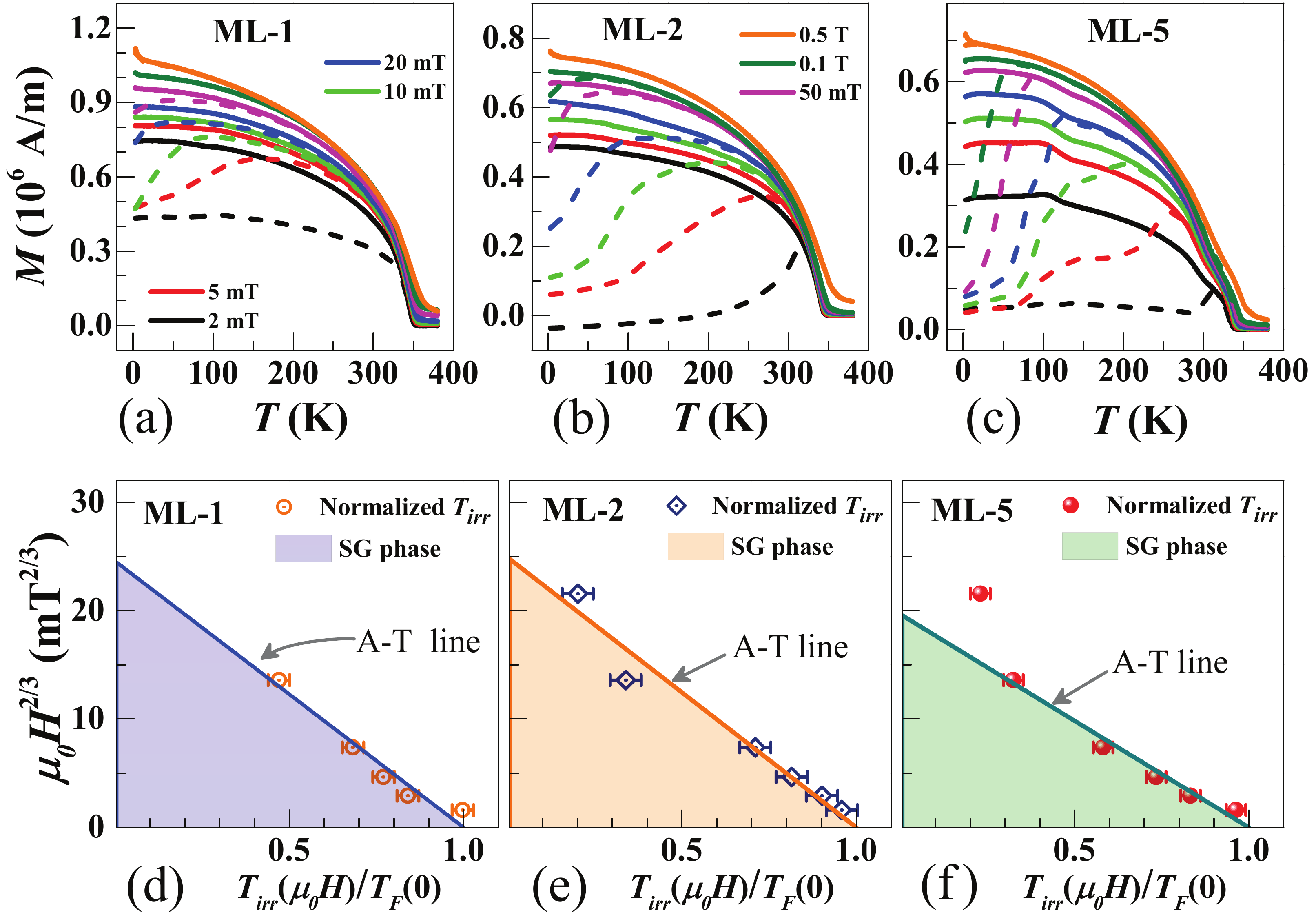}
\caption{\label{Supp_Fig_5_AT_line} $M - T$ variation measured in ZFC-FCW protocol at various magnetic fields for thin films: (a) ML-1, (b) ML-2, (c) ML-5. [Dashed curves: ZFC ; Solid curves FCW]. Plot of de-Almeida-Thouless (A-T) lines showing spin glass (SG) behaviour of the three films: (d) ML-1, (e) ML-2, (f) ML-5. [Symbols represent normalized irreversibility temperatures obtained from experimental $M-T$ curves and the solid lines are fitted A-T lines.]}
\end{figure}

To confirm the presence of spin glass behaviour, magnetization have been measured in ZFC-FCW protocols at various magnetic field strengths for all the films as shown in Fig. S\ref{Supp_Fig_5_AT_line} (a)--(c).

The $T_{irr}$ decreased with the increasing magnetic field and the bifurcation vanished above a certain magnetic field. As shown in Fig. S\ref{Supp_Fig_5_AT_line} (d)-(f), the magnetic field variations of $T_{irr}$ follow de-Almeida Thouless (A-T) line which, within the framework of mean-field theory, is given by:
\begin{equation} \label{Eqn:1}
\frac{\mathit{\mu_{0} H(\mathit{T_{irr}})}}{\Delta \mathit{J}} \propto \left ( 1-\frac{\mathit{T_{irr}}}{\mathit{T_F(0)}} \right )^{3/2}
\end{equation}
where $T_F(0)$ is the spin-glass freezing temperature at $\mu_{0}H$ = 0 mT (actual temperature where spin glass state sets in) and $\Delta J$ is the width of the distribution of the interaction.
However, it should be noted that for all the samples, $T_{irr}$ $<$ $T_{\rm{C}}$. The appearance of an SG transition at a temperature below that of FM ordering characterizes the `re-entrant' spin glass phase.
\subsection{$EB$ loop derivatives}
\begin{figure}[H]
\renewcommand{\figurename}{Figure S}
\centering
\includegraphics[width=\columnwidth]{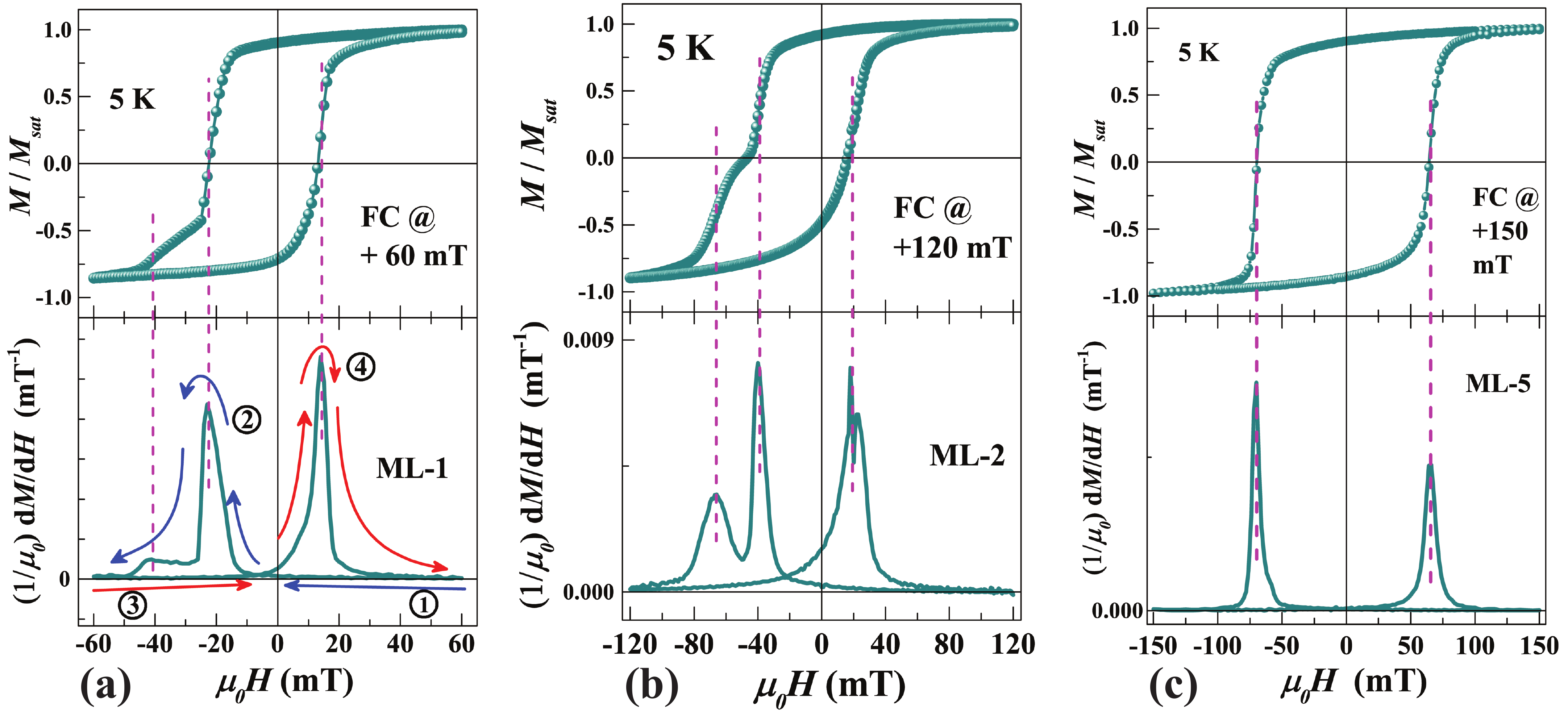}
\caption{\label{Supp_EB_derivative}The $EB$ loops and their field derivatives (taken for positive field cooling cycles) for: (a) ML-1, (b) ML-2 and (c) ML-5.}
\end{figure}
\section*{References}

\end{document}